# Domotic Embedded System

Lidia Dobrescu
" POLITEHNICA" University of Bucharest
ETTI Faculty
Bucharest, Romania
lidia.dobrescu@electronica.pub.ro

*Abstract*— **This paper presents an original domotic embedded system for room temperature monitoring. The OpenRemote is the main software interface between the user and the system, but other software components and communication protocols are used, such as 1-Wire protocol for temperature monitoring devices, RS-232 for the central PC unit and OWFS software for remote control using Android mobile devices. The system architecture consists in hardware and software components to remote control a room temperature parameter for energy efficiency increasing.**

I. INTRODUCTION

The hardware system includes an ATmega16 microcontroller from Atmel manufacturer and a DS18S20 temperature monitoring device with a DS2480B driver as an interface circuit for temperature control.

A central PC with Linux (Ubuntu) operating system is also integrated in the system.

The mobile device can be a smart phone or a common tablet using an Android or IOS operating system.

In the original proposed system, the 1-Wire protocol is used for long distances communications, combining it with OpenRemote free software in an original system controlling the serial port of an Ubuntu server, a mobile device and a microcontroller hardware system. The OpenRemote is free source software working with Java language.

The original 1-Wire protocol was a device communications bus system designed for small networks, on short distances, providing low-speed data and signaling, typically used to communicate with small inexpensive devices such as digital sensors for room temperature or light, with many advantages regarding simplicity and costs. It is extended in this paper beyond its original designed limits

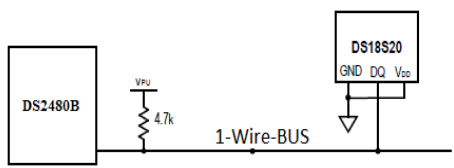

Fig. 1. Temperature sensor and driver

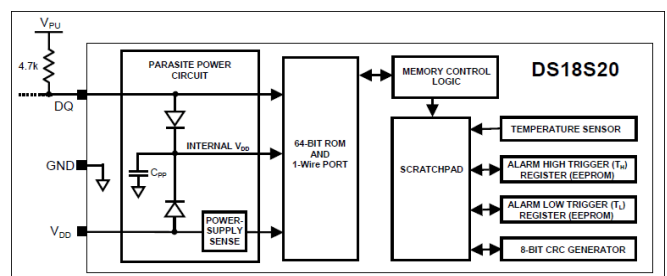

Fig. 2. Block diagram DS18S20 [2]

RS-232 standard and 1-Wire communication protocols are also used in the system. The conversion from RS-232 to TTL voltage signals are realized using MAX232 electronic circuit.

All the above components are embedded using a C program controlling the microcontroller and OpenRemote Design with OWFS software. The OpenRemote program has also been originally interfaced with ATmega16 microcontroller.

The paper describes the new proposed system from the digital thermometer to the command smart phone, highlighting the strong interdependence between hardware and software design.

A real domotic system involves the home automation, in order to keep the constance of the desired monitoring parameters from a room or a building. The proposed system performs the remote temperature monitoring and it can be extended towards a real domotic system for changing the room conditions in order to establish the desired parameter with future mechanical or electrical systems such as heating, ventilation, lighting, shading house appliance.

II. THE TEMPERATURE MONITORING DEVICE

The temperature monitoring is realised using a programmed sensor network [1] and 1-Wire protocol.

The sensor is a DS18S20 integrated circuit and it is connected with DS2480B interface as shown in figure 1.

In fact the sensor is a digital thermometer. Its performance is proven by large temperature measurement limits and temperature step. It also has an alarm function for overpassing the temperature limits settled down by the user in the operating range between −55°C and +125°C. A greater accuracy of ±0.5° is obtained in the range of −10°C to +85°C, suitable to a normal room temperature towards a a real domotic system [2].

The 1-Wire bus is the data line for the sensor DS18S20 for a proper communication with the central microprocessor.

As a designed protocol for short connection, the 1-Wire protocol is widely used for connecting nearby devices. Soon the 1-Wire protocol has became such popular and its users have succed to extend its networked applications beyond the common sizes. In this manner the networks using such a simple and robust protocol have became a complex arrangement of sensors, data lines, and connections [3].

The maximum size extension and the maximum weight of a 1-Wire based network can be hardly calculated. The insufficient drive current provided by the interface is generally a limitation factor. A suitable interface´s impedance in order to avoid signals reflections and interferation with other network slaves are also needed. The reflections, the cable delay, its own resistance conducting to signal degradation are another important limitations for desired network radius.

Another limitation is the network weight. The quite rapid cable charging and discharging can be sometimes a real problem in increasing the network radius. Usually 200m radius can be designed. Increasing the radius towards 500m implies a carefully design.

As the master device, the Atmel microprocessor identifies and addresses the network devices. Each device has a unique code. This important property enables a practically unlimited number of devices that can be accessed on the central bus [2].

Figure 2 shows a block diagram of the used DS18S20 microprocessor. The unique code of the sensors is stored in a ROM on 64 bits. The monitored output signal from the temperature sensor is stored in a 2-byte temperature register. Another two important registers called TH and TL are used for higher and lower threshold alarm trigger. For data storing when the device can be powered down a nonvolatile memory is also used.

The DS18S20 circuit has the ability to operate either with or without an external power supply.

The DS18S20 microprocessor has a VDD power supply. Without this classical power supply it can also operate due to a parasitic power supply extremely useful in in case of space constraints remote applications. In figure 2 this special solution of power supply is indicated using a power capacitor that stores power in high state of the bus.

The 1-Wire network was developped by Dallas Semiconductor as a MicroLan.

The DS2480B microprocessor is the master of 1-Wire network. Its communications lines TXD and RXD from the serial communication port are linked with the 1-Wire bus.

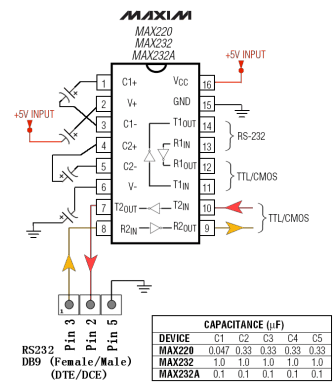

Fig. 4. Pinout diagram DS2480 [4]

The inherent problems related to data rate and speed communication between the port and the 1-Wire bus are solved with configurable parameters.

The diagram in figure 3 shows timing converting circuitry. Different bytes for data, commands and calibrating the time generator are used [4].

All 1-Wire bus and commonly processor data lines speed communication software interfaces are controlled by timing generator.

III. RS -232, TTL CONVERSION AND THE MICROCONTROLLER

RS232 is a commonly used asynchronous serial protocol. It can be used to create a data link between a microcontroller and a standard PC [5]. A microcontroller uses the universal standard called Universal Synchronous Asynchronous Receiver/Transmitter, USART. In 5V system it is a serial data communication The RS232. is used by personal computers. a A conversion between -/+10V of RS232 to to +/- 5V of TTL is required in order to permit a proper communication between the microcontroller and the computer. The MAX232, shown in figure 4 is an integrated circuit for the desired matching in a simple manner shown in figure 5. MAX232 is a voltage converter between RS232 and TTL. As a simple converter with a charge pump for voltage doubling [6].

It includes an electrolytic capacitor. The complete 1-Wire circuit is shown in Figure 5

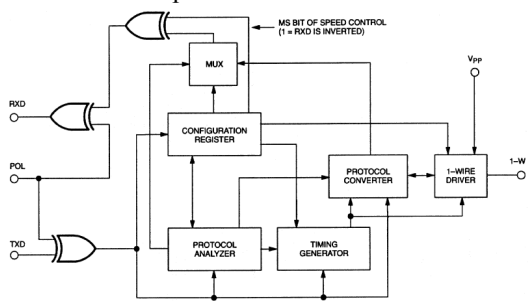

Fig. 3. Block diagram DS2480 [4]

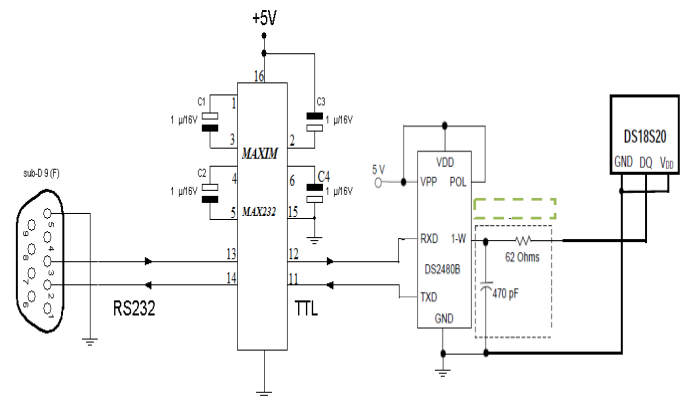

Fig. 5. 1-Wire Circuit

```
int main (void)
{
    DDRD |= 1<<PD6;  //output pins
    DDRD |= 1<<PD4;

    uint8_t press_key;

    ioinit();

    while(1)
    {
        press_key = uart_receive-character ();
    //red LED
    if(press_key == 'b')
        {printf("ON\n");
        PORTD |= (1<< PD4);}
    if(press_key == 'z')
        {printf("OFF\n");
        PORTD &= ~(1<<PD4);}
    if((press_key == 'Y') && (PORTD &= (1<<PD4)))
        printf("on\n");
        else if((press_key == 'Y') && (!(PORTD &= (1<<PD4))))
        printf("off \n");

    //green LED
    if(press_key == 'a')
        {printf("ON\n");
        PORTD |= (1<< PD6);}
    if(press_key == 's')
        {printf("OFF\n");
        PORTD &= ~(1<<PD6);}
    if((press_key == 'X') && (PORTD &= (1<<PD6)))
        printf("ON\n");
        else if((press_key == 'X') && (!(PORTD &= (1<<PD6))))
        printf("OFF \n");

    }
    return(0);
}
```

Fig. 6.    Microcontroller programming

The ATmega16 is a 8-bit AVR RISC, low-power CMOS microcontroller and faster than a CISC one.

Power consumption versus processing speed is carefully optimized [7].

The AVR microcontroller was connected with PC. The data is transmitted from the controller using RS232 standard and displayed on the PC using Hyper Terminal. The USART block is sometimes called the Serial Communications Interface or SCI. The program used for USART was Atmel Studio 6 shown in Figure 6.

In the main program PD6 and PD4 have been defined as the pins where two LEDs, a green and a red one were connected as output loads. A new variable press_key was defined to keep the input character. An init function for the serial link was defined. An infinite while was created for input character. The „b" character for lighting and „z" for stop. „Y"is for interrogation.

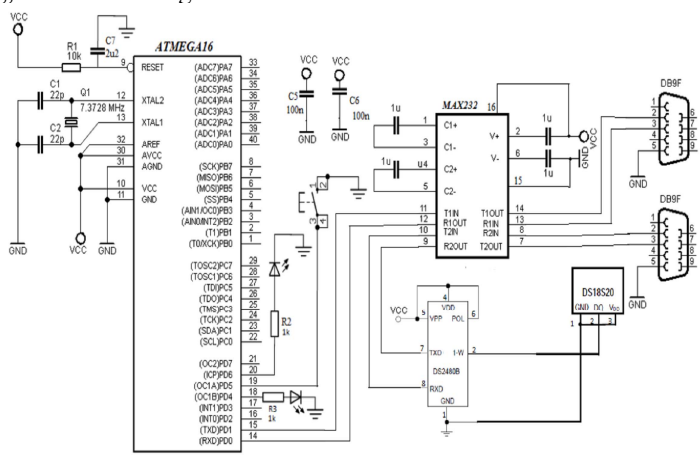

Fig. 7.   The functional circuit

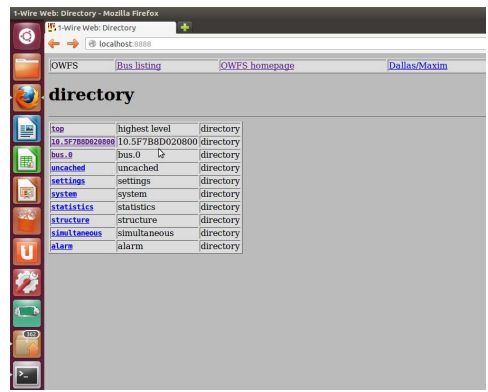

Fig. 8.    Webserver *Owhttpd*

The microcontroller will return „on" or „off" depending on the LED state. The complete circuit is shown in figure 7.

It contains a power-on reset circuit, a LM 7805 regulator and two LEDs. The first green LED lights when the microcontroller PD6 pin has a logic one.

## IV. OWFS PROGRAMS

OWFS consists in a set of programs that can easily access the 1-wire. A virtual file system is created using the same ID as the directory and the individual properties of the device files [8]. OWhttpd is designed as a web server. The figure 8 shows the main screen for this server.

The *temphigh* or *templow* are temperature thresholds, and their overpassing trigger the alarm state in the alarm directory.

A non-volatile memory keeps this temperature thresholds in no power state.

It is preceded by the connection of the network 1-Wire system through the serial port to the PC with the operating system Linux version Ubuntu 12.04.

A command terminal is started to descend into the root privileged mode on. After the install of OWFS (which can perform by "Ubuntu Software Center" simple) more OWFS lines are shown in figure 9.

## V. OPENREMOTE APPLICATION

OpenRemote is an integrated platform. It brings together several residential automation protocols and vendor-specific device protocols and allows an automation solution building that covers a wide spectrum of devices. 1-wire can be used for identification and authentication, sensing and controlling for example lights or temperature through relays.

```
apt-get install fuse-utils libfuse-dev

chmod 777 /dev/ttyS0

/opt/owfs/bin/owserver -device=/dev/ttyS0 -s 4304

/opt/owfs/bin/owfs -s 4304 /var/1-wire

                                    mkdir /var/1-wire)

/opt/owfs/bin/owhttpd -s localhost:4304 -p 8888

cat /var/1-wire/10.5F7B8D020800/temperature
```

Fig. 9.   OWFS command lines

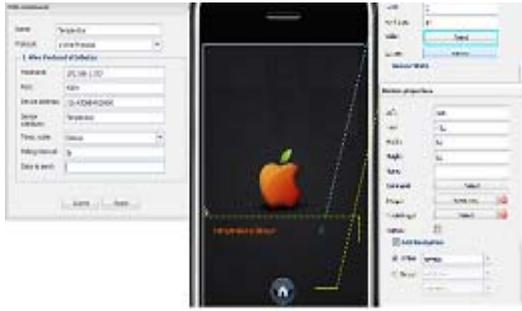

Fig. 10. OpenRemote screen interface

Implementation of 1-Wire support in OpenRemote allows read and write access to 1-Wire network. If a parameter value can be read from 1-Wire device via OWFS, it can be accessed via sensor in OpenRemote. Write commands are executed via buttons, switches or sliders. This will, for example, allow users to control external system via relays.

The software package is free to be downloaded and it is designed for LINUX operating system.

Certified products are extensively tested in combination with Professional Designer and supported by OpenRemote certified Integrators.

Because OpenRemote uses owserver to communicate with physical 1-wire devices, this part must be configured first in the OWFS environment.

Most typical usage pattern of OWFS is to use it as a filesystem. This is very useful access method to 1-wire that provide easy way to test all the devices[8].

With OpenRemote a custom user interface can be created to control the sensor devices. This can be done with OpenRemote Designer and its basic functions of the UI Designer. The application is split into two main sections: the building modeler and the user interface designer.

For many Android, web-based or iOS control devices control panel can be created using user interface design view. The same device can be created in different user profiles.

On each screen several widgets together make up the interface. A widget is a single element that can be interacted with by the user. Each widget has its own functionality and appearance that can be easily modified using a few settings. First, a general overview of the use of widgets in the UI Designer is given [9].

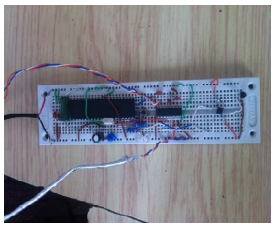

Fig. 11. Hardware circuit

OpenRemote's architecture can be fundamentally configured and changed by user imagination.

„User interface design, installation management and configuration can be handled remotely with OpenRemote cloud-based design tools" [9].

## VI. CONCLUSIONS

The proposed embedded system is designed for energy efficiency increasing and green communications and it allows a remote control of a room temperature for an end-user with a mobile device as a smart phone or a tablet. The sensor output is 1-Wire sent to a microcontroller where it is analised.

A central PC with an installed OWFS server is used as a programmed relay between a mobile device and a microcontroller, as the core of the entire system. The OpenRemote program has been originally interfaced with ATmega16 microcontroller.

The hardware and software solutions are originally putted together using different conversion solutions in a unique architecture.

This project can be the starting point towards a real domotic system as a future house automation.

OpenRemote is a suitable solution to automate modern homes and control heating, TV, lights, media centers using Android devices. It is an open source program control and panel software in a domotic space and it can be integrated with various software protocols.